\documentclass[aps,superscriptaddress,prb,amsmath,amssymb,twocolumn,a4paper,floatfix]{revtex4-1}
\usepackage{graphicx}
\usepackage{float}
\usepackage{hyperref}
\usepackage{bm}
\usepackage{tabularx}
\usepackage{xcolor}
\usepackage{bbm}
\usepackage{accents}
\usepackage{cleveref}

\newcolumntype{Z}{>{\centering\let\newline\\\arraybackslash\hspace{0pt}}X}

\begin{document}
\title{Multi-valuedness of the Luttinger-Ward functional in the Fermionic and Bosonic System with Replicas}
\author{Aaram J. Kim}
\email{aaram.kim@kcl.ac.uk}
\affiliation{Department of Physics, King's College London, Strand, London WC2R 2LS, UK}
\author{Vincent Sacksteder IV}
\affiliation{Department of Physics, Royal Holloway University of London, Egham, Surrey TW20 0EX, UK}

\begin{abstract}
	We study the properties of the Luttinger-Ward functional (LWF) in a simplified Hubbard-type model without time or spatial dimensions, but with $N$ identical replicas located on a single site. 
	The simplicity of this $(0+0)d$ model permits an exact solution  for all $N$ and for both bosonic and fermionic statistics.  
	We show that fermionic statistics are directly linked to the fact that multiple values of the noninteracting Green function $G_0$ map to the same value of the interacting Green function $G$, i.e. the mapping $G_0 \mapsto G$ is non-injective.  
	This implies that with fermionic statistics the $(0+0)d$ model has a multiply-valued LWF.  
	The number of LWF values in the fermionic model  increases proportionally to the number of replicas $N$, while in the bosonic model the LWF has a single value regardless of $N$. 	 
	We also discuss the formal connection between the $(0+0)d$ model  and the $(0+1)d$ model which was used in previous studies of LWF multivaluedness.
\end{abstract}

\maketitle

\section{Introduction}
The Luttinger-Ward functional (LWF)\cite{1960PhRv..118.1417L,Luttinger:1960ch} is the foundation of several modern quantum many-body techniques.
Dynamical mean-field theory (DMFT)\cite{Georges:1996zz} and its extensions\cite{Maier:2005tj,Rubtsov:2008cs,PhysRevLett.107.256402,Pollet:2011jr,Staar:2013ec,Gukelberger:2017en} are derived from the LWF, and have played an important role in studies of both idealized models and real materials~\cite{Kotliar:2006fl,1997JPCM....9.7359A}.
Also based on the LWF are self-consistent diagrammatic methods such as the bold diagrammatic Monte Carlo method with different degrees of the dressing\cite{Prokofev:2007gy,Prokofev:2008jz,VanHoucke:2008ul,Kozik:2010fl}, as well as the self-consistent Hartree-Fock and $GW$ methods~\cite{Biermann:2003hp}.

Despite extensive usage of the LWF, rigorous tests of its formal validity have not been completed.  
The justification of the LWF is based on performing a Legendre transformation on the thermodynamic potential $\Omega$ with respect to the external single-particle source field~\cite{DeDominicis:1964bl,Chitra:2001fp,Potthoff:2003kv,Potthoff:2006hn}. 
Following Baym and Kadanoff, $\Omega$ becomes a stationary value of the transformed functional of the interacting Green function $G$~\cite{Baym:1961do,Baym:1962ci}.
The LWF is the universal part of Baym-Kadanoff functional, which doesn't depend on the noninteracting part of systems.
However the Legendre transformation used to define the LWF depends sensitively on the mathematical properties of the thermodynamic potential $\Omega$.  
If $\Omega$ is not both smooth and convex with respect to the external source field, then the Legendre transformation is not well defined, and the LWF's validity is thrown into doubt.  

A recent study demonstrated the LWF's fragility, showing that when the LWF is applied to a simple model of a fermionic Hubbard atom with $0$ spatial dimensions plus the time dimension (i.e. $(0+1)$ dimensions), then it is a multivalued functional of $G$~\cite{2015PhRvL.114o6402K}. 
In other words, if $G$ is held fixed then two or more values of the LWF and of the self-energy $\Sigma$ can be found which are consistent with the fixed value of $G$.  
The ($(0+1)d$) Hubbard atom has a unique physical value of the self-energy $\Sigma$, so the additional values produced by the LWF are unphysical.

A multiply-valued LWF destroys the predictive power of formalisms such as DMFT and bold diagrammatic Monte Carlo which are based on the LWF to solve interacting systems.  
These formalisms proceed to solution via an iterative process.  
The first step of each iteration is to use the self-energy $\Sigma$  to calculate $G$ via the Dyson equation in combination with the physical $G_0$~.
Then, completing the same iteration, $G$ is used as input to a solver which calculates a new value of $\Sigma$, which will serve as the input of the next iteration.  
It is this final step of the iteration that breaks down in models where the LWF is multivalued, i.e. where $G$ is consistent with several different competing choices of $\Sigma$~.
Given a particular value of $G$, a typical solver will typically find only one of the competing $\Sigma$, and depending on its details it may easily choose one of the non-physical pairs.  
Even if one were able to obtain a complete list of all the competing $\Sigma$, at each iteration of the self-consistent calculation additional input would be required to choose the physical solution.
In order to retain predictive power, the LWF must be singly-valued, and every value of $\Sigma$ must be consistent with a unique value of the interacting Green function $G$.  

The first report that the LWF is multivalued when applied to the $(0+1)d$ fermionic Hubbard atom used  the bold diagrammatic Monte Carlo method and  DMFT  ~\cite{2015PhRvL.114o6402K}. 
Following this watershed paper, several authors have performed detailed studies of the LWF's problem of multiple values.   In order to obtain a  qualitative understanding Ref. ~\cite{Rossi:2015cx}~ introduced a simpler fermionic toy model where time was supressed (a $(0+0)d$ model), and explained the qualitative behavior of the first unphysical branch. 
Ref. ~\cite{Stan:2015en,Tarantino:2017dr} extensively investigated the functional space of the Green function.  Ref. ~\cite{Gunnarsson:2017co} found  additional unphysical branches, implying that the LWF has an infinite number of values when applied to the $(0+1)d$ model, and also found that one eigenvalue of the   charge vertex diverges at the branching point of the LWF~\cite{Schafer:2013gv,Gunnarsson:2017co,Dave:2013hy}.

In spite of these extensive studies, the problem of the LWF's multivaluedness is still not thoroughly understood. It is not clear how general this problem is, or what are the essential ingredients of the set of models for which the LWF produces multiple solutions.
Most importantly, the role of the fermionic statistics in the LWF has not been studied.

In this present paper we study the LWF's behavior when applied to a $(0+0)d$ fermionic model which has been generalized to include $N$ replicas on the single site.  
We also study a model which is different in only one respect: the $N$ replicas obey bosonic statistics instead of fermionic statistics.    
We exactly solve both models, and we find a mathematical correspondence between the fermionic $N$-replica model and the bosonic $N$-replica model: the Green function $G$ of the bosonic $N$-replica model is exactly the same as the Green function of the fermionic $N$-replica model with $-N$ substituted for $N$.   
This means that the bosonic results can be obtained from the fermionic results, and vice versa, by changing the sign of the replica count $N$. 
With these results in hand, we examine the number of possible values of the LWF as a function of $N$, for strictly real $G$ and also for complex $G$.  
In the fermionic model the number of values increases with $N$ with a staircase profile, while in contrast the bosonic model always has exactly one solution.
In other words, in the $(0+0)d$ model the multivaluedness of the LWF is caused specifically by fermionic statistics, and is cured by using bosonic statistics. 
Examining the mathematical structure of the model, we find that the sign of the fermionic partition function $\mathcal{Z}$ changes as the single-particle potential is varied, and that at each sign change the thermodynamic potential $\Omega= -\ln \mathcal{Z}$ is not smooth. 
In addition, $\Omega$ is in  general not convex.  
These two properties result in the multiple values.  
In contrast, in the bosonic case $\mathcal{Z}$ has a single sign and $\Omega$ is both smooth and convex, resulting in a single-valued LWF.    

\section{Model and Method}
We introduce the $(0+0)d$ Hubbard model with $N$ replicas, a single particle potential $\mu$, and a quartic interaction with strength $U$. 
The actions $\mathcal{S}_F$ and $\mathcal{S}_B$ of the fermionic and bosonic variants are:
\begin{eqnarray}
	\label{eqn:SF}
	\mathcal{S}_F &=& \frac{U}{2}\left(\sum^{N}_{\alpha=1}\sum^{}_{\sigma}\bar{\psi}_{\alpha\sigma}\psi_{\alpha\sigma}\right)^2 - \mu\sum^{N}_{\alpha=1}\sum^{}_{\sigma}\bar{\psi}_{\alpha\sigma}\psi_{\alpha\sigma}~,\\
	\label{eqn:SB}
	\mathcal{S}_B &=& \frac{U}{2}\left(\sum^{N}_{\alpha=1}\sum^{}_{\sigma}\bar{\varphi}_{\alpha\sigma}\varphi_{\alpha\sigma}\right)^2 - \mu\sum^{N}_{\alpha=1}\sum^{}_{\sigma}\bar{\varphi}_{\alpha\sigma}\varphi_{\alpha\sigma}~. 
\end{eqnarray}
Here   $\psi_{\alpha\sigma},\, \bar{\psi}_{\alpha\sigma}$ are  fermionic Grassmann variables with the replica index $\alpha = 1, \cdots, N$ and  spin $\sigma$, and $\varphi_{\alpha\sigma},\,\bar{\varphi}_{\alpha\sigma}$ are  complex  bosonic variables.  
Note that there is no imaginary-time index in Eqn.~(\ref{eqn:SF},\ref{eqn:SB}), implying that there is no Hamiltonian-based description of the $(0+0)d$ model.
The suppression of the imaginary-time index can also be regarded as the high temperature limit of $(0+1)d$ theory.
This simplifies the functional space of the LWF: the Green function is a single number, and the bare Green function $G_0$ is equal to $G_0 = \mu^{-1}$.   
These conveniences allow us to easily investigate the analytic structure of its mapping from $G_0$ to $G$.

The same model occurs in the replica theory of random matrices, with $U$ being the disorder strength, and $\mu$ controlling which eigenvalues one is investigating\cite{Kamenev:1999bg}. In that setting physical results are obtained by using the ``replica trick", which involves treating $N$ as a continuous not integer variable and taking the $N\rightarrow 0$ limit.      In contrast, here we have an exploratory focus, and are interested in the LWF's behavior for all $N$.

We calculate the fermionic partition function using the combinatorics of the Grassmann variables:
\begin{eqnarray}
	\mathcal{Z}_F(N,U,\mu) &=& \int_{}^{}\prod_{\alpha=1}^{N}\prod_{\sigma=\uparrow,\downarrow}d\bar{\psi}_{\alpha\sigma}d\psi_{\alpha\sigma}\exp\left\{ -\mathcal{S}_F \right\}~\\
	&=& \sum^{N}_{n=0}\frac{1}{n!}\int_{}^{}\prod_{\alpha=1}^{N}\prod_{\sigma=\uparrow,\downarrow}d\bar{\psi}_{\alpha\sigma}d\psi_{\alpha\sigma} (-\mathcal{S}_F )^n~\\
	&=& \sum^{N}_{k=0}\frac{(2N)!}{(2N-2k)!(k!)}\left(-\frac{U}{2}\right)^k\mu^{2(N-k)}~\\
	&=& \left(2U\right)^N \mathcal{U}\left(-N,\frac{1}{2},\frac{\mu^2}{2U}\right)~.
	\label{}
\end{eqnarray}

On the third line we used the binomial expansion of $(-\mathcal{S}_F )^n$.  
The final result is written in terms of $\mathcal{U}$, the Tricomi confluent hypergeometric function, which is defined for both integer and non-integer $N$.  
In the case of integer values of the number of replicas $N$ the partition function $\mathcal{Z}_F$ is an $N$th order polynomial in $\mu^2$ and $U$.  
It therefore is able to change sign as a function of $\mu$ up to $2N$ times,  and it never diverges for any finite value of $\mu^2$ and $U$.

In contrast, the bosonic partition function is always positive.  Moreover it   converges only if either $U>0$, or if $U=0$ and $\mu < 0$:
\begin{widetext}
\begin{eqnarray}
	\mathcal{Z}_B(N,U,\mu) &=& \int_{}^{}\prod_{\alpha=1}^{N}\prod_{\sigma=\uparrow,\downarrow}\left[d\bar{\varphi}_{\alpha\sigma}d\varphi_{\alpha\sigma}\right]\exp\left\{ -\mathcal{S}_B \right\}~\\
	&=& \int_{}^{}\prod_{\alpha=1}^{N}\prod_{\sigma=\uparrow,\downarrow}\left[d\bar{\varphi}_{\alpha\sigma}d\varphi_{\alpha\sigma}\right]\exp\left\{ -\frac{U}{2}\left( \sum^{}_{\alpha\sigma}\bar{\varphi}_{\alpha\sigma}\varphi_{\alpha\sigma} \right)^2 + \mu \sum^{}_{\alpha\sigma}\bar{\varphi}_{\sigma\alpha}\varphi_{\alpha\sigma}\right\}~\\
	&=& \frac{\Omega_{4N}}{2}\int_{0}^{\infty}dQ~Q^{2N-1}\exp\left\{ -\frac{U}{2}Q^2 + \mu Q\right\}~\\
	&=& \frac{1}{2}\left(\frac{2\pi^2}{U}\right)^{N}\left[\frac{\Gamma\left(\frac{1}{2}\right)}{\Gamma\left(\frac{1}{2}+N\right)}\mathcal{M}\left(N,\frac{1}{2},\frac{\mu^2}{2U}\right)+\left(\frac{-\mu}{\sqrt{2U}}\right)\frac{\Gamma\left(-\frac{1}{2}\right)}{\Gamma\left(N\right)}\mathcal{M}\left(\frac{1}{2}+N,\frac{3}{2},\frac{\mu^2}{2U}\right)\right]~,\\
	&=& \frac{1}{2}\left(\frac{2\pi^2}{U}\right)^{N}\mathcal{U}\left(N,\frac{1}{2},\frac{\mu^2}{2U}\right)~.
	\label{}
\end{eqnarray}
\end{widetext}
$\mathcal{M}$ is the Kummer confluent hypergeometric function and $\Omega_{4N}=2\pi^{2N}/\Gamma(2N)$ is the solid angle of the $4N$-dimensional sphere.  Comparison of the final result for the bosonic $\mathcal{Z}_B(N,U,\mu)$ with the fermionic $\mathcal{Z}_F(N,U,\mu) = \left(2U\right)^N \mathcal{U}\left(-N,\frac{1}{2},\frac{\mu^2}{2U}\right)$ shows that, up to a normalization constant that is independent of $\mu$, the bosonic and fermionic results are the same Tricomi confluent hypergeometric function $\mathcal{U}$, with the only difference being $N \rightarrow -N$.

It is worth noting that the $N \rightarrow -N$ correspondence between bosonic and fermionic results is well known in the literature of replicas.  This correspondence is natural because a Gaussian integral with commuting variables produces an inverse determinant, while a Gaussian integral with Grassmann variables produces a determinant.  It is also common to treat the number of replicas $N$ as a real variable; this is the foundation of the replica approach where $N$ is analytically continued to $N=0$.  

Next we calculate the Green function:
\begin{eqnarray}
	\label{eqn:GF}
	G_{F}&=&\frac{1}{2N}\frac{\partial}{\partial\mu}\log\mathcal{Z}_{F}~\nonumber\\
	&=& \frac{\mu}{2U}\frac{\mathcal{U}\left(1-N,3/2,\mu^2/2U\right)}{\mathcal{U}\left(-N,1/2,\mu^2/2U\right)}~\\
	\label{eqn:GB}
	G_{B}&=&-\frac{1}{2N}\frac{\partial}{\partial\mu}\log\mathcal{Z}_{B}~\nonumber\\
	&=& \frac{\mu}{2U}\frac{\mathcal{U}\left(1+N,3/2,\mu^2/2U\right)}{\mathcal{U}\left(N,1/2,\mu^2/2U\right)}~.
	\label{}
\end{eqnarray}
Here we find an exact correspondence between bosonic and fermionic results under the transformation $N \rightarrow - N$. 
There is however an immense difference between the bosonic $+N$ case and the fermionic $-N$ case: Since the bosonic partition function $\mathcal{Z}_{B}$ is positive definite and is finite (if $U>0$), the bosonic Green function has no isolated pole.  
In contrast, the sign of the fermionic partition function $\mathcal{Z}_{F}$ can change as  many as $2N$ times, each of which causes a pole in the fermionic Green function.

In order to analyze the number of branches of the LWF, we express the interacting  Green function $G$ as a function of the noninteracting Green function $G_0 = \mu^{-1}$:
\begin{equation}
	G\left[N,U;G_0\right] = \frac{1}{2G_0 U}\frac{\mathcal{U}\left(1+N,3/2,1/2G_0^2U\right)}{\mathcal{U}\left(N,1/2,1/2G_0^2 U\right)}~.
	\label{}
\end{equation}
The LWF is free from the problem of multiple values if for every value of $G$ there is only one value of $G_0$ which produces that value.

\section{Results}
\begin{figure}[]
	\centering
	\includegraphics[width=0.35\textwidth]{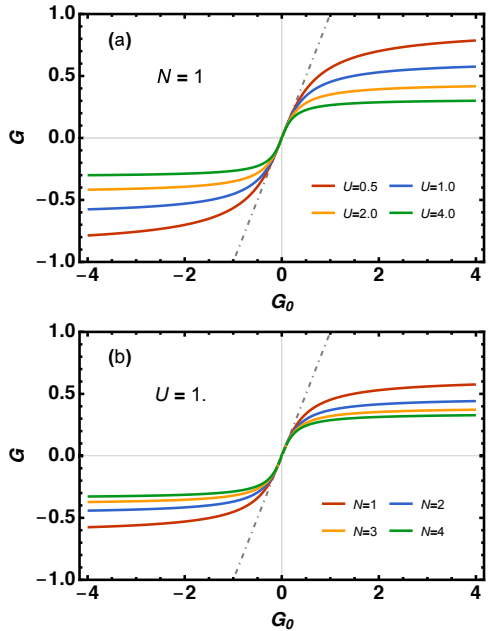}
	\caption{	
		The interacting Green function $G\left[G_0\right]$ in the bosonic $(0+0)d$ model.  
		$G$ increases monotonically with $G_0$, and for each value of $G$ there is a unique value of $G_0$, showing that the LWF is single-valued.  
		In panel (a), the number of replicas $N=1$ and the interaction strength $U = 0.5$ (red line, largest values), $1.0$ (blue), $2.0$ (yellow), and $4.0$ (green, smallest values).  
		In panel (b), the interaction strength $U=1.0$ and the number of replicas $N = 1$ (red line, largest values), $2$ (blue), $3$ (yellow), and $4$ (green, smallest values).  
		The dot-dashed line shows $G=G_0$.
	}
	\label{fig:boson}
\end{figure}
In Figure~\ref{fig:boson} we plot the  bosonic $G\left[N,U;G_0\right]$ for various $U$ and $N$ values.
The map $G_0\mapsto G$ is always injective for all positive $N$ and $U$ values investigated.
The injective map implies the well-defined Legendre transformation of the thermodynamic potential.
In fact, the thermodynamic potential is a convex function with respect to $\mu$ of a fixed sign.
So the Legendre transformation for a given sign of $\mu$ is well-defined.
Furthermore, since $\mu$-negative and -positive (or equivalently positive and negative $G_0$) branch map to $G$ of different signs, respectively, the injective mapping from $G_0\mapsto G$ is preserved for all sign of $G_0$~.

\begin{figure*}[]
	\centering
	\includegraphics[width=0.9\textwidth]{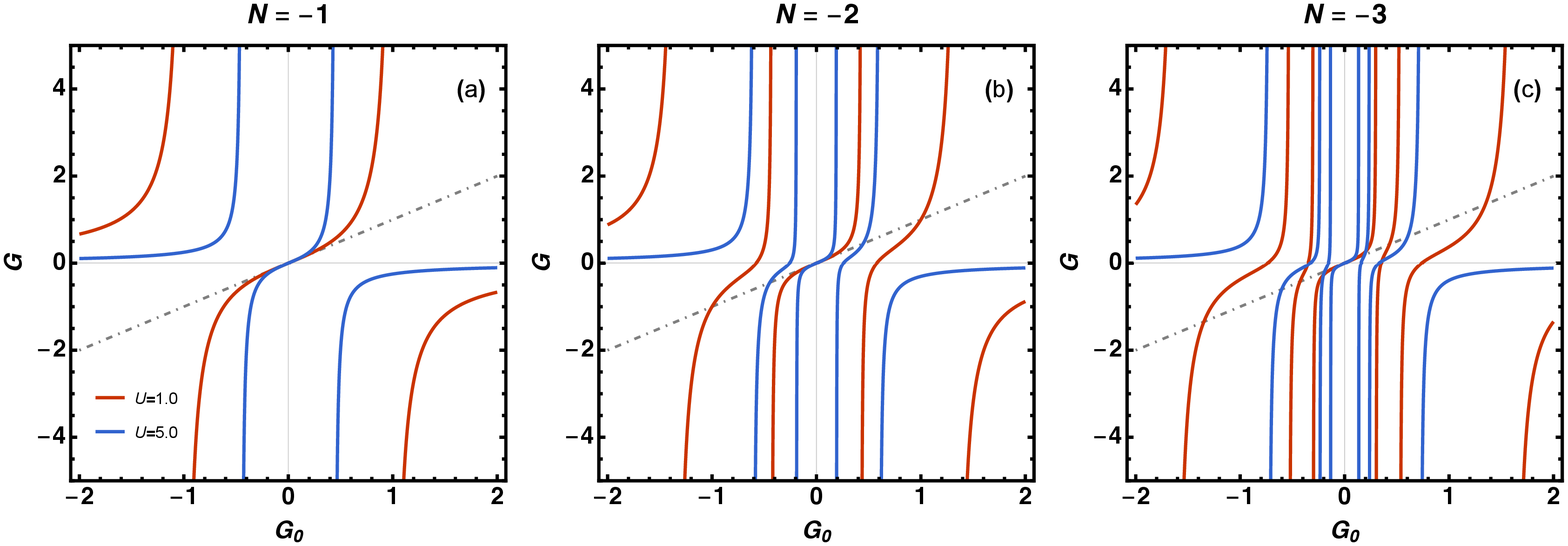}
	\caption{
		The interacting Green function $G[G_0]$ in the fermionic $(0+0)d$ model.   
		In panel (a), the model has $1$ replica, $G$ has  $2$ poles, and for a given value of $\tilde{G}$ there are $2$ solutions $\tilde{G}_0$ which satisfy $G[\tilde{G}_0] = \tilde{G}$.
		In panel (b), there are $2$ replicas and $4$ poles,  and for each value of $G$ there are $4$ solutions of $G_0$.  
		In panel (c), there are $3$ replicas, $6$ poles, and $6$ solutions of $G_0$.   
		Red lines show $G$ with an interaction strength $U=1.0$ and blue lines show $U=5.0$. 
		The dot-dashed lines show $G=G_0$.
	}
	\label{fig:fermion}
\end{figure*}
The well-defined LWF in the bosonic case is in contrast to the fermionic one where the map is not injective.
Figure~\ref{fig:fermion} shows the fermionic $G_0\mapsto G$ mapping for three different integer $N$ values; $N=-1$, $-2$, and $-3$~.
For $N=-1$, there exist one positive and one negative $\tilde{G}_0$ for a given $\tilde{G}$, which satisfy $G[\tilde{G}_0]=\tilde{G}$~.
The number of the both positive and negative solution increases by $1$ as we decrease $N$ by $1$, respectively.
So total increase in the number of the solutions is $2$ for an additional replica index.
As we decrease $N$ by $1$, the number of poles along the real axis increases by $2$, which is the same increase of the number of solutions.
And poles along the real axis correspond to the sign change of the partition function.

\begin{figure}[]
	\centering
	\includegraphics[width=0.35\textwidth]{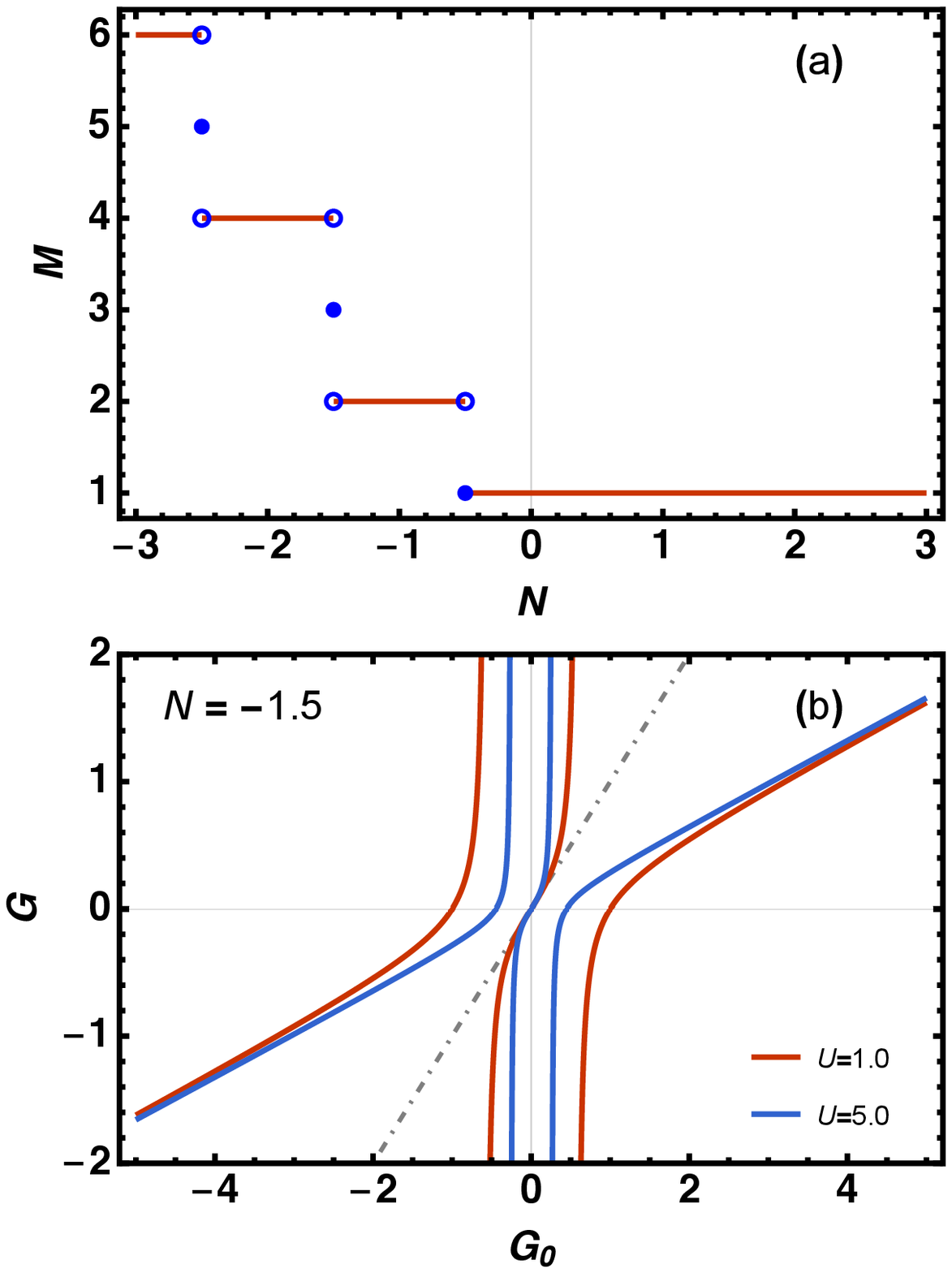}
	\caption{
		The number $M$ of values of $G_0$ that give the same value of $G$, as  function of the number of replicas $N$, in the fermionic $(0+0)d$ model.  
		Negative values of $N$ on this graph report results from the fermionic model with $|N|$ replicas, while positive values report the behavior of the bosonic model.  
		$N$ is treated as a continuous variable because the fermionic and bosonic partition functions are a hypergeometric function that is defined for all real $N$.  
		At negative half integers the number of solutions $M$ value is odd, and is represented as solid blue dots.
		Panel (b) shows the case of $N=-3/2$, where there are $3$ values of $G_0$ that give the same value of $G$.  
		$G$ diverges proportionally to $G_0$ at large $|G_0| \gg 1$. 
		Red lines show $G$  with an interaction strength $U=1.0$ and blue lines show $U=5.0$. The dot-dashed line shows $G=G_0$.
	}
	\label{fig:Msolution}
\end{figure}
The evolution of the number of the solutions $M$ as a function of $N$ is shown in Fig.~\ref{fig:Msolution}(a).
In Fig.~\ref{fig:Msolution}(a), the number of replica indices is generalized to the real number instead of the integer.
Clear step-like increase in $M$ is observed only in the fermionic side with the negative $N$.
On the $N$-axis, the discontinuous change in $M$ occurs at the negative half-integer $N$; $-1/2$, $-3/2$, $-5/2$, and so on.
The $N=-1/2$ case corresponds to the spinless system with vanishing interacting term since the total number of indices becomes unity and the self-interaction term becomes quadratic due to the fermionic statistics.
So the interacting Green function $G$ becomes the same as the noninteracting one $G_0$.
For the negative half-integer $N$ smaller than $-1/2$, there appears the most outer branch $G$ branch whose range spans $\left(-\infty,\infty\right)$ instead of $\left(0,\infty\right)$/$\left(-\infty,0\right)$, giving additional $G_0$ for a given $G$.
Figure~\ref{fig:Msolution}(b) presents the $N=-3/2$ case whose $M=3$ for $U=1$, and $5$~.

\begin{figure*}[]
	\centering
	\includegraphics[width=0.80\textwidth]{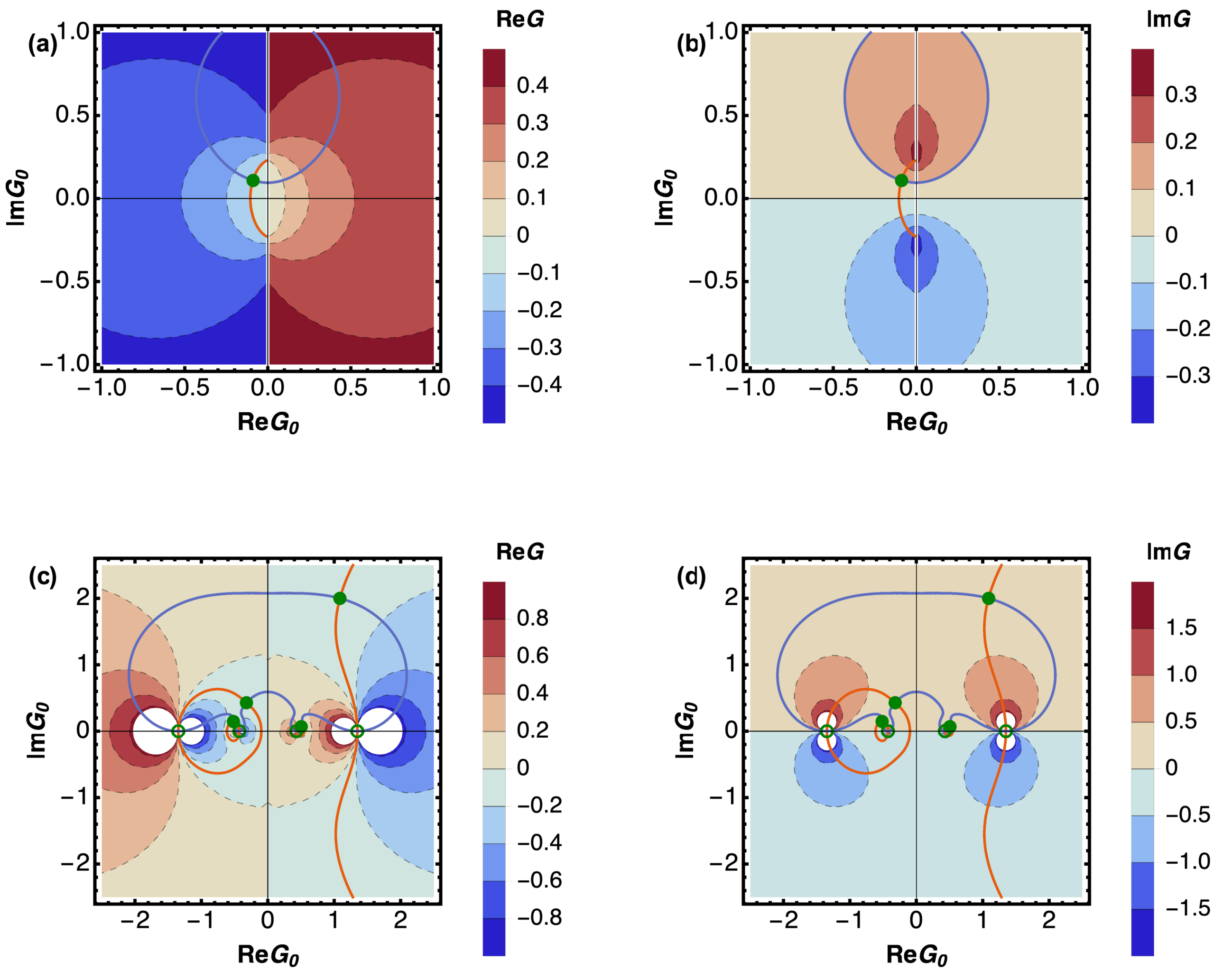}
	\caption{
		$G[G_0]$ when both $G$ and $G_0$ are complex, with filled green dots placed at the solutions of $G[\tilde{G}_0]=\tilde{G}$. 
		Contours of $\mathrm{Re}G$ are shown on the left (panels a,c) and contours of $\mathrm{Im}G$ are shown on the right (panels b,d). 
		The bosonic $G$ is shown in the upper panels (a,b), with a red line at $\mathrm{Re}G = -0.1$, a blue line at $\mathrm{Im}G = 0.1$, and a filled green dot at the solution of $G = \tilde{G} = -0.1+0.1i$.   
		The fermionic $G$ is shown in the lower panels (c,d), with red lines at $\mathrm{Re}G = -0.1$,  blue lines at $\mathrm{Im}G = 0.35$, and four filled green dots at the four  solutions of $G = \tilde{G} = -0.1+0.35i$.  
		The open green dots are located at singularities of $G$ and do not solve  $G = \tilde{G}$. 
		The number of replicas is $|N|=2$ in all panels and the interaction strength $U=1.0$.
	}
	\label{fig:complexG0}
\end{figure*}
We now generalize the domain of the mapping $G_0\mapsto G$ to complex number.
Figure~\ref{fig:complexG0} shows the contour plot of the real and imaginary part of $G$ on the complex $G_0$ plane for both bosonic ($N=2$) and fermionic ($N=-2$) case, respectively.
For a given $\tilde{G}$ the $\mathrm{Re}\tilde{G}$ contour is highlighted with a red line, and the $\mathrm{Im}\tilde{G}$ contour is highlighted with a blue line.  
The solution $\tilde{G}_0$ which satisfies $G[\tilde{G}_0]=\tilde{G}$ appears as an intersection of two contour lines for real and imaginary part of $\tilde{G}$~.

As we gradually introduce the imaginary part to $\tilde{G}$, the solutions $\tilde{G}_0$ evolve to general complex number from real number.
In general, the number of solutions is preserved in a presence of the imaginary part of $\tilde{G}$~.
One can find the single intersection for the bosonic case ($N=2$) marked as (green) solid circle, but four different solutions for the fermionic case ($N=-2$).
Note that four additional intersections are marked with open circles for the fermionic case, but these are not true solutions because they lie at singularities where $\mathrm{Re}G$ changes discontinuously from $+\infty$ to $-\infty$.

\begin{figure}[]
	\centering
	\includegraphics[width=0.35\textwidth]{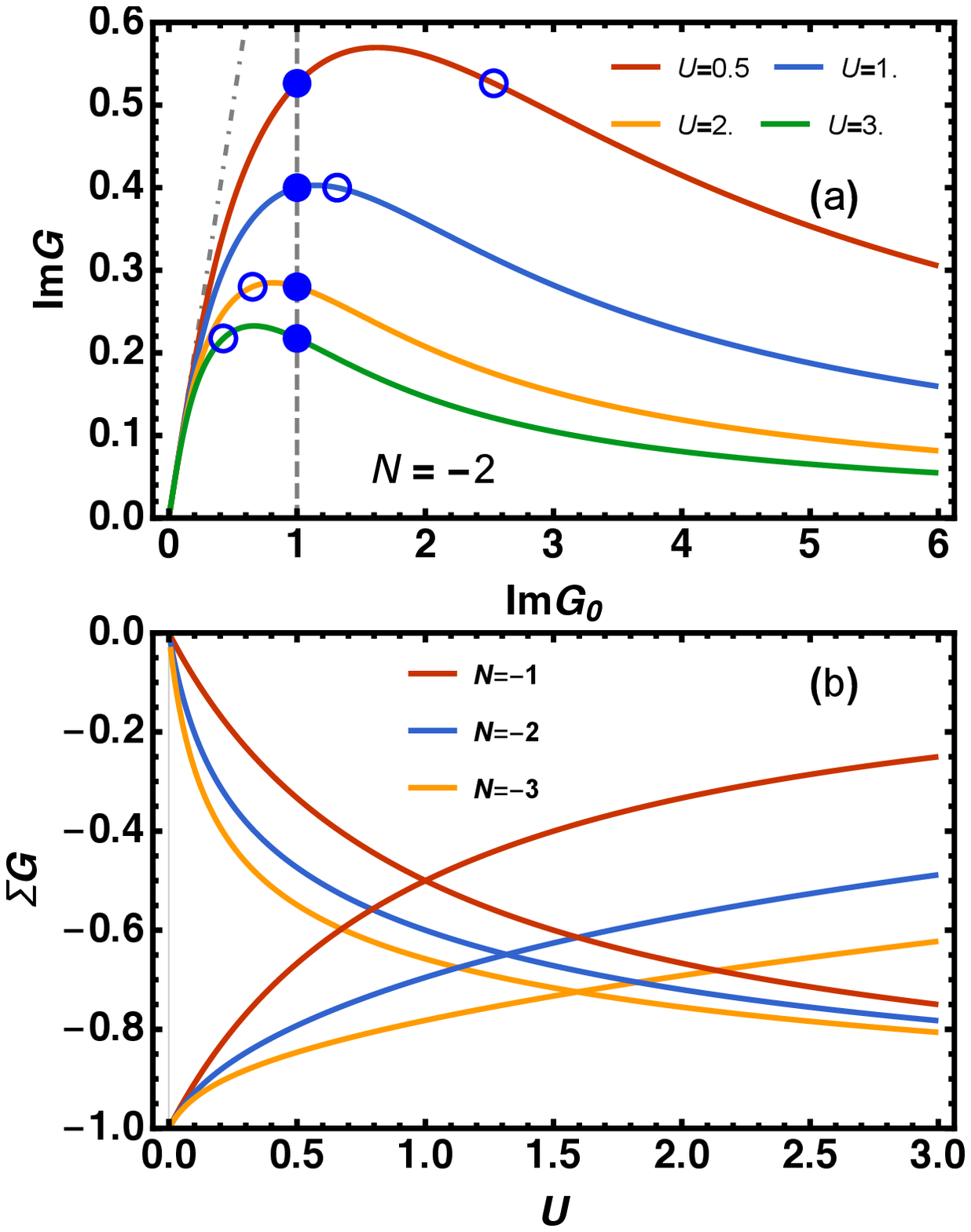}
	\caption{
		(a) Fermionic interacting Green function as a function of purely imaginary noninteracting Green function for $N=-2$~.
		Red, blue, orange, and green line represent $U=0.5$, $1.0$, $2.0$, and $3.0$, respectively.
		Blue solid dots show the physical solutions $(\tilde{G}_{0,phys},\tilde{G}_{phys})$ for a given imaginary chemical potential $\mu=-i$ with various $U$ values.
		Blue open symbols presents the unphysical solutions $(\tilde{G}_{0,unphys},\tilde{G}_{phys})$.
		As we increase $U$, $\tilde{G}_{0,unphys}$ crosses the $\tilde{G}_{0,phys}$ at $U_c\sim 1.3$~.
		Dash-dotted line shows $G=G_0$~.
		(b) Evolution of the $\Sigma[\tilde{G}_{phys}]\tilde{G}_{phys}$ for two different $\tilde{G}_{0}$ branches.
		The physical branches are connected to the noninteracting limit, $\Sigma G=0$~.
		There exist a branching point between the physical and the unphysical branches for each $N$ value and it becomes larger as we increase $|N|$ values.
	}
	\label{fig:ImG0}
\end{figure}
We also present a special case where two different $\tilde{G}_0$ solutions can be degenerate in the fermionic case.
Supposing a purely imaginary $\mu$ which results in the purely imaginary physical noninteracting Green function $\tilde{G}_{0,phys}=1/\mu$, the physical interacting Green function $\tilde{G}_{phys}$ is also purely imaginary.
However, there exists an additional unphysical solution $\tilde{G}_{0,unphys}$ which is purely imaginary.
As we increase the $U$ values,  $\tilde{G}_{0,unphys}$ approaches toward $\tilde{G}_{0,phys}$ from higher absolute values, and eventually become degenerate at $U_c$~.
For $U>U_c$, the absolute value of $\tilde{G}_{0,unphys}$ becomes smaller than $\tilde{G}_{0,phys}$~.

Figure~\ref{fig:ImG0}(a) shows the evolution of $\tilde{G}_{0,phys}$ and $\tilde{G}_{0,unphys}$ as a function of $U$ for $N=-2$ fermionic case.
One can find the crossing of $\tilde{G}_{0,phys}$ and $\tilde{G}_{0,unphys}$ at $U_c\sim 1.3$~.
Figure~\ref{fig:ImG0}(b) presents the $\Sigma[\tilde{G}_{phys}]\tilde{G}_{phys}$ of two different branches as a function of $U$~.
Here, the self-energy is defined as a function of $G$, $\Sigma[G] = 1/G_{0}[G] - 1/G$~.
Since $G_{0}[\tilde{G}_{phys}]$ has two different branches, $\Sigma[\tilde{G}_{phys}]$ also has two corresponding branches.
The branching point $U_c$ increases as we decrease $N$~.
We also emphasize that there exist additional $\tilde{G}_{0,unphys}$ away from the imaginary axis for $N\leq -2$~.
But those additional solutions don't become degenerate for a purely imaginary $\tilde{G}$~.

The origin of the multiple branches along the imaginary axis is the lack of the log-convexity of the partition function.
For a purely imaginary chemical potential $\mu$ there exist two inflection points in the thermodynamic potential as a function of $\mu$~.
Two inflection points separate the domain of the $G_0\mapsto G$ mapping into three pieces, and for a given sign of $\tilde{G}$, two of three $G_0$ domains incorporate $\tilde{G}$ in the corresponding image.
The existence of the physical and the unphysical solutions $\tilde{G}_0$ is the manifestation of such domain structure.

Finally, we comment on the connection between  the $(0+1)d$ Hubbard atom and the model studied here which has $(0+0)d$ and $N$ replicas.  
After introduction of a finite temperature $T = \beta^{-1}$ and decomposition into Matsubara frequencies $\omega_n$, where $n$ ranges from $-\infty$ to $+\infty$, the  action for the $(0+1)d$ fermionic model is 
\begin{eqnarray}
	\mathcal{S}_{HA} &=& -\frac{1}{\beta}\sum^{}_{n\sigma}\bar{\psi}_{n\sigma}(i\omega_n+\mu)\psi_{n\sigma} \nonumber\\
	&&+ \frac{U}{\beta^3}\sum^{}_{nmk}\bar{\psi}_{n-k,\uparrow}\psi_{n,\uparrow}\bar{\psi}_{m+k,\downarrow}\psi_{m,\downarrow}~.
	\label{}
\end{eqnarray}
This action is formally similar to the $(0+0)d$ model,  with the number of replicas taken as equal to the number of Matsubara frequencies, i.e. countably infinite.  
The main differences are simplifications: in the $(0+0)d$ model the single-particle term's frequency dependence is supressed, and the energy transfers in the interaction term are removed.

Despite these simplifications, the $(0+0)d$ model's qualitative behavior shows several remarkable similarities to that of the $(0+1)d$ model.  
First of all, the convergence of the skeleton series to the unphysical branch in the $(0+1)d$ model can be understood in terms of the crossing of the physical and unphysical $G_0$ of the $(0+0)d$ model.
For purely imaginary $\tilde{G}$, two out of $2|N|$ solutions are aligned on the imaginary axis, and cross each other at $U=U_c$~.
Since the skeleton series always chooses the weakly interacting $\Sigma$, the skeleton series converges to the unphysical branch for $U>U_c$~.

The infinite number of solution of the LWF observed in A. Toschi el. al.\cite{Gunnarsson:2017co} clearly appears in our model in the $|N|\rightarrow \infty$ limit.
In the $(0+0)d$ model, the total number of branches scales as $2|N|$ as $|N|\rightarrow \infty$~.
One of the interesting observations is that there exists branches which breaks the structures of the solution for the $(0+1)d$ Hubbard atom.
For examples, A. Toschi et. al. showed that there exist $G_0$s which show the non-trivial real part in contrast to the exact $G_0$ which is purely imaginary.
For the general negative $N$, all solutions but two along the imaginary axis show non-trivial real part for a given purely imaginary $\tilde{G}$~.

Furthermore, our model suggests that the skeleton series for the bosonic case is promising.
Throughout our study, the bosonic model shows the well-defined LWF without the multivaluedness problem.
Our results give the positive signal to the bosonic bold-diagrammatic Monte Carlo method whose major concern is the possible multivaluedness problem of the skeleton series.

\section{Conclusion}
We exactly solve the $(0+0)d$ model with the general number of replicas for both bosons and fermions.
It turns out that both the bosonic and the fermionic Green function can be written in terms of the Tricomi confluent hypergeometric function, but with different sign of the number of replica index, $N$~.
We show that the multivaluedness of the LWF is only observed for fermionic model not bosonic one implying the direct link to the fermionic statistics.
Especially, the sign oscillation and the lack of the $\log$-convexity of the partition function are the characteristic feature of the fermionic statistics in the $(0+0)d$ model.
In the fermionic model, the multiple $G_0$s result in the same $G$, and the number of $G_0$ increases proportional to the number of replicas.
For a complex $G$, the multiple $G_0$s evolve to complex numbers.
We found the interesting case where two purely imaginary $G_0$ can be degenerate, which resembles the unphysical branch of the skeleton series of $(0+1)d$ model.
Furthermore, our results of the simple toy model is a positive signal to the bosonic bold series whose main concern is the convergence to unphysical solutions.

\section{Acknowledgement}
AJK was supported by EPSRC grant EP/P003052/1 and VES was supported by EPSRC grant EP/M011038/1. 
We thank L. Du, A. Toschi, V. Olevano, C. Weber, E. Plekhanov, H. Kim, L. Reining, P. Werner, E. Kozik, and B. Svistunov for fruitful discussions.
VES thanks King’s College London for hospitality.
\bibliography{ref}

\end{document}